\documentclass[11pt,a4paper]{article}
\usepackage{fullpage}
\usepackage{xcolor,graphicx,tikz}
\usetikzlibrary{arrows.meta}
\usepackage{amssymb,amsmath,amsthm}
\usepackage[colorlinks, citecolor=blue, linkcolor=blue!50!black, urlcolor=blue!50!black]{hyperref}
\hypersetup{pdftitle={A note on input-oblivious unambiguous hierarchies}, pdfauthor={Edward A. Hirsch}}
\newcommand{\poly}{\mathrm{poly}}
\newcommand{\lex}{<_{\ell ex}}
\newcommand{\gtlex}{>_{\ell ex}}
\newcommand{\gelex}{\ge_{\ell ex}}
\newcommand{\bit}{\{0,1\}}

\newcommand{\Out}{\mathrm{Out}}
\newcommand{\Acc}{\mathrm{Acc}}
\newcommand{\Good}{\mathrm{Good}}
\newcommand{\class}[1]{\ensuremath{\mathbf{{#1}}}}
\renewcommand{\P}{\class{P}}
\newcommand{\FP}{\class{FP}}
\newcommand{\pr}{\class{pr}}
\newcommand{\co}{\class{co}\textbf{-}}
\newcommand{\NP}{\class{NP}}
\newcommand{\UP}{\class{UP}}
\newcommand{\ONP}{\class{ONP}}
\newcommand{\OUP}{\class{OUP}}
\newcommand{\SOUP}{\class{SOUP}}
\newcommand{\RP}{\class{RP}}
\newcommand{\BPP}{\class{BPP}}
\newcommand{\ZPP}{\class{ZPP}}
\newcommand{\MA}{\class{MA}}
\newcommand{\AM}{\class{AM}}
\newcommand{\OMA}{\class{OMA}}
\newcommand{\PP}{\class{PP}}
\newcommand{\SPP}{\class{SPP}}

\newcommand{\OP}{\class{\oplus P}}
\newcommand{\UAP}{\class{UAP}}
\newcommand{\PH}{\class{PH}}
\newcommand{\OH}{\class{OH}}

\newcommand{\OIP}{\class{OIP}}
\newcommand{\YP}{\class{YP}}
\newcommand{\TALLY}{\class{TALLY}}
\newcommand{\SPARSE}{\class{SPARSE}}
\newcommand{\OptP}{\class{OptP}}

\newcommand{\Ppoly}{\class{P/poly}}

\newcommand{\OMIN}{\class{OMIN}}
\newcommand{\LtP}{\class{L_2^{\class{\scriptscriptstyle P}}}}
\newcommand{\OLtP}{\class{OL_2^{\class{\scriptscriptstyle P}}}}
\newcommand{\StP}{\class{S_2^{\class{\scriptscriptstyle P}}}}
\newcommand{\OtP}{\class{O_2^{\class{\scriptscriptstyle P}}}}
\newcommand{\SigmakP}[1]{\class{\Sigma_{#1}^{\class{\scriptscriptstyle P}}}}
\newcommand{\OGTD}{\class{OGTD}}
\newcommand{\OUGTD}{\class{OUGTD}}
\newcommand{\OM}{\class{OM}}
\newcommand{\OUM}{\class{OUM}}
\newcommand{\sOprMA}{\class{sOprMA}}
\newcommand{\UH}{\class{UH}}
\newcommand{\OUH}{\class{OUH}}
\newcommand{\GUH}{\class{\mathcal{UH}}}
\newcommand{\AUPH}{\class{AUPH}}
\newcommand{\PUH}{\class{UH_\bullet}}
\newcommand{\POUH}{\class{OUH_\bullet}}

\newcommand{\POUHK}{\class{OU\Sigma_\bullet}}

\newcommand{\OUSigmaP}[1]{\class{OU\Sigma_{#1}^{\class{\scriptscriptstyle P}}}}

\makeatletter
\newcommand*\bigcdot{\mathpalette\bigcdot@{.5}}
\newcommand*\bigcdot@[2]{\mathbin{\vcenter{\hbox{\scalebox{#2}{$\m@th#1\bullet$}}}}}
\makeatother
\renewcommand{\dot}{\bigcdot}
\newcommand{\dexists}{\exists\dot}
\newcommand{\doexists}{\exists_{\class{o}}\dot}
\newcommand{\duexists}{\exists_{\class{u}}\dot}

\newcommand{\douexists}{\exists_{\class{ou}}\dot}
\newcommand{\douforall}{\forall_{\class{ou}}\dot}
\newcommand{\BP}{\class{BP}\dot}
\DeclareFontShapeChangeRule{it}{sc}{sc}{sc}
\newcommand{\problem}[1]{\ensuremath{\textsc{#1}}}

\newcommand{\YES}{\problem{\tiny YES}}
\newcommand{\NO}{\problem{\tiny NO}}
\numberwithin{equation}{section}
\newtheorem{theorem}{Theorem}[section]
\newtheorem{lemma}[theorem]{Lemma}
\newtheorem{proposition}[theorem]{Proposition}
\newtheorem{corollary}[theorem]{Corollary}
\newtheorem{definition}[theorem]{Definition}
\newtheorem{remark}[theorem]{Remark}
\newtheorem*{nonumtheorem}{Theorem}
\newtheorem*{nonumlemma}{Lemma}
\newcommand{\etalchar}[1]{$^{#1}$}

\usepackage[normalem]{ulem}

\title{A note on input-oblivious unambiguous hierarchies}
\author{Edward A. Hirsch\thanks{Department of Computer Science, Ariel University, Israel.
Email: {\tt edwardh@ariel.ac.il}. This research was conducted with the support of the State of Israel, the Ministry of Immigrant Absorption, and the Center for the Absorption of Scientists.}}
\date{November 21, 2026}

\begin{document}
\maketitle

\begin{abstract}
Unambiguous hierarchies are defined similarly to the polynomial hierarchy; however, all witnesses must be unique. Their versions differ in the mode of using oracles. We consider input-oblivious unambiguous hierarchies built from the class $\OUP$, in which all positive inputs of the same length have the same unique witness: the hierarchy $\OUH$ with language oracles and the hierarchy $\POUH$ with loose access to promise oracles, which is the input-oblivious version of the loose unambiguous hierarchy of Hirsch and Sokolov (IPL, 2015). We show that languages in $\POUH$ can be decided in deterministic polynomial time with an $\ONP$ oracle, and for languages in $\OUH$ oracle queries can be made nonadaptive.

Moreover, languages in $\OUH$ can be decided in deterministic polynomial time with a tally $\NP$ oracle, and languages in $\POUH$ can be decided in $\P/\OptP$, that is, a subclass of $\Ppoly$ where the advice is generated by a function that maximizes outputs of a nondeterministic machine on a unary input. We show that $\P/\OptP$ can be characterized in terms of $\OGTD$ and $\OUGTD$ generator--tester--decoder classes recently considered by Hirsch and Volkovich (2026) and, in particular, is contained in their ``ultimate'' Karp--Lipton collapse class $\OUM$.

We also observe several other facts related to $\OUP$: (1) the hierarchy defined by input-oblivious unambiguous dot-operators collapses, (2) in $\OUP$ computations, a single loose query to  $\pr\OUP$ can be replaced by a query to  $\OUP$, (3) $\SOUP=\OUP\cap\co\OUP$, where $\SOUP$ is the same-witness class recently considered by Hirsch, Volkovich, and Williams (2026).
\end{abstract}

{\small

\paragraph{AI usage statement.}
All notions and statements in this paper were formulated by the author, who also shared his proof sketches and discussed the proofs with the AI tools GPT6-Astra and Claude Opus~5.5. After that Opus~5.5 assisted with the literature search (the author admits that as a result he read the amount of \emph{classical}
literature like never before) and with drafting the text. The final version was written by the author in a long dialogue with Opus~5.5; in particular, the author verified and edited all parts written by the above tools and takes full responsibility for the content of the paper.

}

\section{Introduction}

\paragraph{Unambiguous hierarchies.}
Nondeterministic polynomial-time computations with at most one accepting path constitute the class $\UP$ defined\footnote{Here and in what follows, for the definitions of classical complexity classes, we refer the reader to the Complexity Zoo (\href{https://complexityzoo.net}{\nolinkurl{complexityzoo.net}}) or to \cite[Appendix~A]{HO02}, or (for the most basic ones) to \cite{AB09}.} by Valiant. 
Unambiguous hierarchies~\cite{LR94,NR98} are defined similarly to the polynomial hierarchy; however, all witnesses must be unique. These hierarchies have subtle differences in the mode of using oracles. The hierarchy $\UH=\UP\cup\UP^{\UP}\cup\UP^{\UP^{\UP}}\cup\ldots$~\cite{NR98} uses oracle access to languages, and the computation needs to be unambiguous only for the correct oracle. The hierarchy $\GUH$~\cite{LR94,NR98} uses \emph{guarded} oracle access to promise problems~\cite{CHV93}: queries outside the promise set are prohibited (Grollmann and Selman~\cite{GS88} called such reductions \emph{smart}). The unambiguous alternation hierarchy $\AUPH$~\cite{LR94} is defined by unambiguous alternating machines with a bounded number of alternations, via single-query oracle access, or by alternating unambiguous dot-operators~\cite{NR98} (Niedermeier and Rossmanith attribute the equivalence to Hemaspaandra). 
These hierarchies satisfy $\AUPH\subseteq\UH\subseteq\GUH$ levelwise and $\GUH\subseteq\SPP$~\cite{LR94};
moreover, $\UH$ and $\GUH$ are contained in the unambiguous alternating polynomial-time class $\UAP$~\cite{CGRS04}, which is contained in $\SPP$~\cite{NR98}. Hemaspaandra and Rothe~\cite{HR97} related their structure to the existence of sparse Turing-complete sets for $\UP$, and Spakowski and Tripathi~\cite{ST07,ST09} studied their relativized structure.

Hirsch and Sokolov~\cite{HS15} relaxed the definition of oracle access and considered the \emph{loose} unambiguous hierarchy $\pr\PUH$: queries outside the promise set are allowed, but the computation must be correct (and unambiguous) for \emph{all} possible answers to such queries. This mode of access to promise problems was introduced for deterministic machines by Cai, Hemachandra, and Vysko\v{c}~\cite{CHV93} under the name ``arbitrary failure access''; it was used by Chakaravarthy and Roy~\cite{CR11} for deterministic computations querying $\pr\MA$ and $\pr\AM$, and it is also meant when the Valiant--Vazirani theorem (see~\cite[Theorem~17.18]{AB09}) is formulated as $\NP\subseteq\RP^{\pr\UP}$.
The class $\PUH$ (the class of languages in $\pr\PUH$) contains $\UH$ and $\GUH$ and is still contained in $\PH$, levelwise. Hirsch and Sokolov strengthened the first part of Toda's theorem $\PH\subseteq\BP\OP\subseteq\P^{\PP}$ (see~\cite[Section~17.4]{AB09}) to $\PH=\BP\pr\PUH$. %
They asked whether $\pr\PUH\subseteq\pr\SPP$~\cite[Section~4]{HS15},
which has been very recently answered affirmatively by 
Miloschewsky, Podder, and Rudolph~\cite[Lemma~3.6]{MPR26}.

\paragraph{Input-oblivious computations.}
In an input-oblivious computation, the witness may depend on the length of the input but not on the input itself. Chakaravarthy and Roy~\cite{CR06} defined the input-oblivious version $\OtP$ of $\StP$, Fortnow, Santhanam, and Williams~\cite{FSW09} defined the input-oblivious versions $\ONP$ and $\OMA$ of $\NP$ and $\MA$, and Goldreich and Meir~\cite{GM15} studied input-oblivious proof systems as a uniform perspective on $\Ppoly$. These classes are contained in $\Ppoly$; for example, $\NP\subseteq\Ppoly$ if and only if $\NP=\ONP$~\cite[Proposition~5]{FSW09}. Hirsch, Volkovich, and Williams~\cite{HVW26} recently introduced the class $\OUP$ of input-oblivious unambiguous computations, where all positive inputs of the same length have the same unique witness, and the class $\OMIN$, which equals the input-oblivious version $\OLtP$~\cite{HV26a} of the class $\LtP$ of Korten and Pitassi~\cite{KP24}. They proved $\OUP\subseteq\OMIN$ and considered the oblivious hierarchy $\OH=\ONP\cup\ONP^{\ONP}\cup\ldots\subseteq\OtP$. It is natural to consider input-oblivious unambiguous hierarchies as well (and this is what we do in this paper): the hierarchy $\OUH=\OUP\cup\OUP^{\OUP}\cup\ldots$ with language oracles and the loose hierarchy $\pr\POUH$; by $\POUH$ we denote the class of languages in $\pr\POUH$.

\paragraph{Several-stage computations.}
Hirsch and Volkovich~\cite{HV26b} introduced generator--tester--decoder computations. On inputs of length $n$, a \emph{generator}, which is a polynomial-time transducer with an oracle, computes an advice string from $1^n$; a \emph{tester} certifies the advice without seeing the input; and a polynomial-time \emph{decoder} decides the input using the advice. The class $\OGTD[\mathcal A,\mathcal B,\mathcal C]$ consists of the languages decided in this way by a generator from $\mathcal A$, a tester from $\mathcal B$, and a decoder from $\mathcal C$; in $\OUGTD[\mathcal A,\mathcal B,\mathcal C]$ the tester accepts exactly one advice string for each $n$ (see Section~\ref{subsec:ogtd}). An important particular case, denoted $\OGTD[\mathcal X]$, is when the generator is in $\FP^{\mathcal X}$, the decoder is in $\P$, and no tester is required. %
Clearly, $\OGTD[\mathcal X]\subseteq\P^{\mathcal X}$, since a polynomial-time machine with an oracle from $\mathcal X$ can compute the advice and then run the decoder.

Two-stage computations of this kind have been studied for decades. Karp and Lipton defined advice classes $\mathcal C/f$
that allowed computational models of class $\mathcal C$ to use the same trusted (and potentially uncomputable) advice string of length $f(n)$ on all inputs of length~$n$.
In particular, $\Ppoly=\OGTD[\poly,\ast,\P]$, where $\poly$ stands for arbitrary polynomially bounded advice (a potentially uncomputable function $1^n\mapsto a_n$) and $\ast$ means there is no tester. Ko and Sch\"oning~\cite{KS85}, K\"obler and Thierauf~\cite{KT94}, and Gavald\`a~\cite{Gav95} bounded the complexity of computing correct advice, that is, the complexity of the generator, and K\"amper's nonuniform proof systems~\cite{Kam91} delimit the complexity of the sets of correct advice, that is, the sets accepted by a tester; for the usual classes $\mathcal B$ and $\mathcal C$, $\OGTD[\poly,\mathcal B,\mathcal C]$ is his class $\mathcal C\text{-}\mathcal B\text{-}\class{PSL}$.
K\"obler and Thierauf used the notation $\mathcal C/\mathcal F$ to require the advice strings to be computed by functions $1^n\mapsto a_n$ in a complexity class $\mathcal F$. In their notation,
$\OGTD[A]=\P/\FP^A$ for a language $A$. In fact, $\OGTD[\NP]=\P/\OptP$ (see Corollary~\ref{cor:popt}, which also follows from Krentel's theorem~\cite[Theorem~3.1]{Kre88}); here $\OptP$ is Krentel's class of optimization functions. 

Aaronson~\cite{Aar07} introduced the class $\YP$ of languages decidable in polynomial time with untrusted advice, and Aaronson and Drucker~\cite{AD14} introduced its variant $\YP^*$, where the advice is tested in polynomial time without access to the input; $\YP^*$ is $\OGTD[\poly,\P,\P]$,
that is, K\"amper's class $\P\text{-}\P\text{-}\class{PSL}$~\cite{Kam91}. Input-oblivious proof systems~\cite{FSW09,GM15} can be viewed as restricting $\Ppoly$ to verifiable advice; in the class $\OIP$ of Goldreich and Meir~\cite{GM15}, the interaction takes place on $1^n$, and its result is subsequently used on the actual input. 
Buhrman, Fortnow, Hitchcock, and Loff~\cite[Theorem~4]{BFHL13} showed that if $\NP\subseteq\Ppoly$ and $\PH\subseteq\P^{\NP}$, then $\PH\subseteq\P/\alpha$ for an advice function~$\alpha$ whose graph is in $\co\NP$, that is, with unique advice certified by a $\co\NP$ tester. The main class of~\cite{HV26b} is $\OM=\OGTD[\FP^{\sOprMA},\co\ONP,\P]$, where $\sOprMA$ is a restricted version of $\pr\OMA$ (still containing $\ONP$). They also proved that a Karp--Lipton collapse holds down to its unique-advice subclass $\OUM$ (that is, $\NP\subseteq\Ppoly$ implies $\PH=\OUM$), which is contained in three Karp--Lipton collapse classes that are pairwise incomparable as far as is known to date, $\OUM\subseteq\OM\subseteq\OMIN\cap\P^{\sOprMA}\cap\ZPP^{\ONP}$.

\subsection{Our results}

\paragraph{Input-oblivious unambiguous hierarchies.}
It is known that $\OUP\subseteq\OMIN$ and that $\OMIN$ is closed under polynomial-time Turing reductions~\cite{HVW26}; thus one can ask whether $\OUH\subseteq\OMIN$ and what is an upper bound on $\POUH$. We answer the former question affirmatively; even more, we show that the languages of both hierarchies have polynomial-size circuits that can be constructed from $1^n$ with an $\NP$ oracle. For $\OUH$, nonadaptive queries to a tally $\NP$ language suffice:
\begin{nonumtheorem}[Theorem~\ref{th:ouh}]
$\OUH\subseteq\P^{\TALLY\cap\NP}$.
\end{nonumtheorem}
In particular, every language in $\OUH$ can be decided in polynomial time with nonadaptive queries to a language in $\ONP$, i.e., $\OUH\subseteq\P^{\ONP}_\parallel$ (Corollary~\ref{cor:ouh}). 
For loose access to promise oracles, we construct an adaptive generator:
\begin{nonumtheorem}[Theorem~\ref{th:loose}]
$\POUH\subseteq\OGTD[\NP]$.
\end{nonumtheorem}
For generators with an $\NP$ oracle, the advice can be made canonical, which gives a unique advice string certified by a $\co\ONP$ tester:
\begin{nonumlemma}[Lemma~\ref{lem:canonical} and Corollary~\ref{cor:popt}]
$\P/\OptP=\OGTD[\NP]=\OGTD[\ONP]=\OUGTD[\FP^{\ONP},\co\ONP,\P]$.
\end{nonumlemma}
It puts this class (as well as $\POUH$) into both $\P^{\ONP}$ and $\OUM$.
Consequently, every language in $\POUH$ can be decided in polynomial time with adaptive queries to a language in $\ONP$: $\POUH\subseteq\P^{\ONP}$ (Corollary~\ref{cor:loose}). 

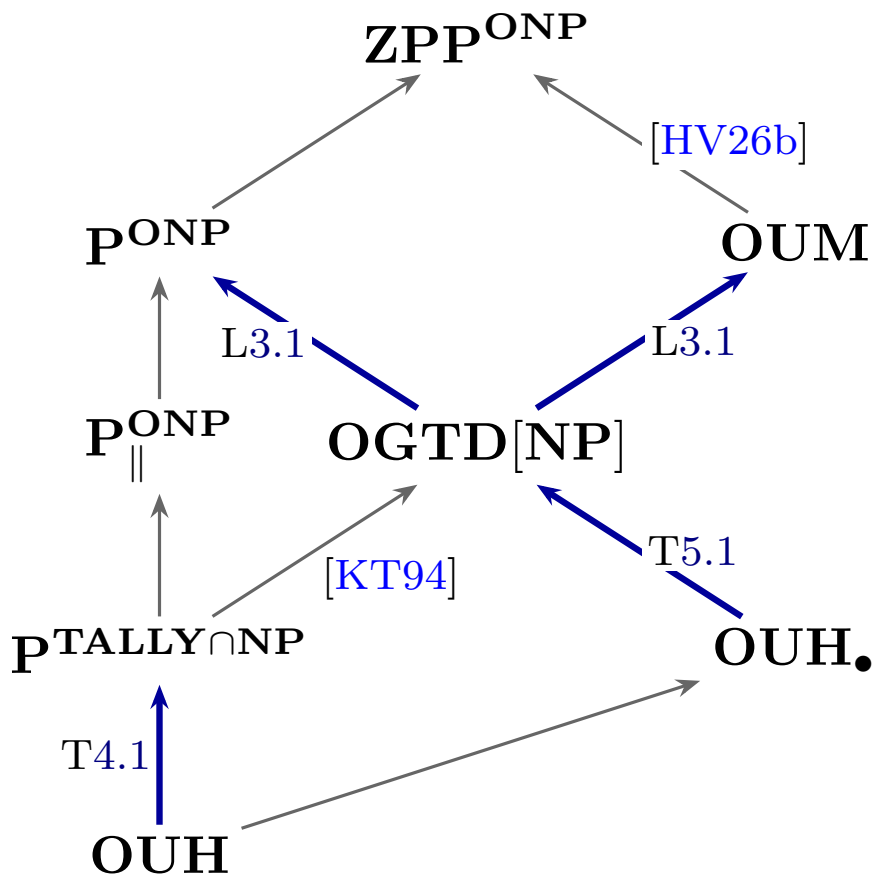
\begin{figure}[t]
\centering
\begin{tikzpicture}[x=1.4cm,y=0.9cm,
  every node/.style={font=\small,inner sep=2pt,align=center},
  inc/.style={-{Stealth[length=1.6mm]},semithick,draw=black!60},
  res/.style={inc,very thick,draw=blue!60!black},
  lab/.style={font=\scriptsize,fill=white,inner sep=1pt}]
\node (ouh) at (0,0) {$\OUH$};
\node (tally) at (0,1.5) {$\P^{\TALLY\cap\NP}$};
\node (pouh) at (3,1.5) {$\POUH$};
\node (ponppar) at (0,3) {$\P^{\ONP}_{\parallel}$};
\node (ogtd) at (1.5,3) {$\OGTD[\NP]$};
\node (ponp) at (0,4.5) {$\P^{\ONP}$};
\node (oum) at (3,4.5) {$\OUM$};
\node (zpp) at (1.5,6) {$\ZPP^{\ONP}$};
\draw[res] (ouh)--(tally) node[midway,left,lab]{T\ref{th:ouh}};
\draw[inc] (ouh)--(pouh);
\draw[inc] (tally)--(ponppar);
\draw[inc] (tally)--(ogtd) node[midway,below right,lab,text width=1.5cm,align=left]{\cite{KT94}};
\draw[res] (pouh)--(ogtd) node[midway,right,lab]{T\ref{th:loose}};
\draw[inc] (ponppar)--(ponp);
\draw[res] (ogtd)--(ponp) node[midway,left,lab]{L\ref{lem:canonical}};
\draw[res] (ogtd)--(oum) node[midway,right,lab]{L\ref{lem:canonical}};
\draw[inc] (ponp)--(zpp);
\draw[inc] (oum)--(zpp) node[midway,right,lab]{\cite{HV26b}};
\end{tikzpicture}
\caption{\itshape Containments between the main classes considered in this paper; arrows point to larger classes. Thick arrows are proved in this paper; the other arrows are known or immediate. All these classes are contained in $\Ppoly$. Moreover, $\OGTD[\NP]=\OUGTD[\FP^{\ONP},\co\ONP,\P]=\P/\OptP$ by Lemma~\ref{lem:canonical} and Corollary~\ref{cor:popt}.}\label{fig:map}
\end{figure}

\medskip
See Fig.~\ref{fig:map} for the containment diagram. 

\paragraph{Dot-operators and single queries.}
\leavevmode One could try to define our hierarchies using dot-operators instead of oracles. Dot-operators were introduced by Wagner~\cite{Wag86} and studied further in~\cite{ZF87,Zac88,Sch89,BGM06}. The polynomial hierarchy can be defined using alternating existential and universal dot-operators (for example, $\NP^{\NP}=\dexists\co\NP$), and so can $\OH$, since the proof of $\ONP^{\ONP}=\doexists\co\ONP$~\cite[Theorem~4.4]{HVW26} obviously relativizes; also the hierarchy $\AUPH$ can be equivalently defined using alternating unique dot-operators~\cite{LR94,NR98}. In contrast, the hierarchy defined by alternating input-oblivious unambiguous dot-operators collapses to $\OUP\cup\co\OUP$ (Theorem~\ref{th:dot} and Corollary~\ref{cor:dot-collapse}). We also observe that a single loose query to a promise problem in $\pr\OUP$ can be replaced by a query to a language in $\OUP$, that is, $\OUP^{\pr\OUP[1]}=\OUP^{\OUP[1]}$, and similarly $\UP^{\pr\UP[1]}=\UP^{\UP[1]}$ (Theorem~\ref{th:one-query}). This is in contrast with $\RP^{\pr\RP[1]}=\BPP$ (by the proof of~\cite[Theorem~1]{BF99}), whereas $\P=\RP\neq\BPP$ relative to an oracle~\cite[Theorem~3]{BF99}, and with $\ONP^{\pr\ONP[1]}=\OtP$~\cite[Theorem~4.7]{HVW26}, whereas $\ONP^{\ONP[1]}=\ONP^{\ONP}$ is not known to equal $\OtP$. Finally, the same-witness class $\SOUP$ of~\cite{HVW26} equals $\OUP\cap\co\OUP$; hence $\OUP^{\SOUP}=\OUP$ and $\SOUP^{\SOUP}=\SOUP$ (Proposition~\ref{prop:soup}).

\subsection{Our methods}\label{subsec:methods}

\paragraph{Canonical advice.}
To better locate the class $\OGTD[\NP]$, where we put our hierarchies, 
we discuss it in Section~\ref{sec:generators}.
Our Lemma~\ref{lem:canonical} is based on Krentel's observation~\cite[Theorem~3.1]{Kre88} that the correct sequence of answers of an $\NP$ oracle is the lexicographically largest sequence in which all positive answers are certified; with some effort, the lemma can also be inferred from his results. Buhrman, Fortnow, Hitchcock, and Loff~\cite[proof of Theorem~4]{BFHL13} obtained unique advice with a $\co\NP$ graph in a similar way, appending to the run of an $\FP^{\NP}$ generator the lexicographically smallest witnesses for its positive answers. We encode a run of the generator on $1^n$ by its sequence of oracle answers followed by certificates for the positive answers, and take the lexicographically largest feasible encoding $a_n$ as the advice. This way, $a_n$ is a common witness for all queries of the prefix search that finds it, so it is found in $\FP^{\ONP}$; it also refutes every other feasible encoding, so it is the only string accepted by a $\co\ONP$ tester. A minimum of a linear order is used as a common refutation in the same way in~\cite{HVW26,HV26b}.

\paragraph{Language oracles.}
For $\OUH$ (Theorem~\ref{th:ouh}) we show that $\P^{\TALLY\cap\NP}$ is closed under $\OUP$ oracle computations. We use Mahaney's census technique (see~\cite[Section~1.2]{HO02}), which was also used for computing advice from $1^n$ in~\cite{KT94} and for unambiguous computations with sparse oracles in~\cite{HR97}: knowing the number of strings of a tally $\NP$ oracle up to the relevant length, an $\NP$ machine can certify all its answers. In this way, K\"obler and Thierauf~\cite[Theorem~3.2]{KT94} proved $\NP^{\SPARSE\cap\NP}\subseteq\NP/\OptP[O(\log n)]$, where the advice is this number of strings. For a tally language in $\NP^{\TALLY\cap\NP}$, this number can be obtained by asking the tally oracle itself, and the remaining question is a query to another tally language in $\NP$; hence every tally language in $\NP^{\TALLY\cap\NP}$ belongs to $\P^{\TALLY\cap\NP}$ (Lemma~\ref{lem:tally}; with some effort, it can also be inferred from the results of~\cite{KT94}). Our new ingredient is the obliviousness of the witness of the outer $\OUP$ computation: since all positive inputs of length~$n$ share the unique witness~$w_n$, its bits are answers to oracle $\NP$ questions that do not depend on the input, and these questions can be answered by %
a tally language in $\NP^{\TALLY\cap\NP}$. 

\paragraph{Loose access.}
For promise oracles, the census technique does not work because of unclear attribution of out-of-promise queries, and a search for an accepting witness may find a witness accepted on an input outside the promise. Our generator for $\POUH$ (Theorem~\ref{th:loose}) follows the strategy of the proof by Hartmanis and Hemachandra~\cite[Theorem~2.1]{HH90} that ``robustly categorical'' machines (nondeterministic machines with at most one accepting path for every oracle) accept, relative to every oracle $A$, languages 
decidable in polynomial time with queries to~$A$ and to an $\NP$ oracle; 
the result was extended to ``robustly unambiguous'' 
alternating machines (machines with a bounded number of alternations in which, for every oracle, every existential configuration has at most one accepting successor and every universal configuration has at most one rejecting successor) in~\cite[Theorem~5.1]{ST07}.
Their simulation repeatedly finds, with an $\NP$ oracle, an accepting path consistent with the answers of $A$ learned so far and asks $A$ all queries of this path; it stops when the path is correct or no such path exists. Since two accepting paths for different oracles must conflict on some query, each new path conflicts with every previous path on a query on which the latter was wrong; hence the number of rounds is bounded by the number of queries on a path. Spakowski and Tripathi~\cite{ST07} apply this argument level by level: by induction on the number~$k$ of levels, they simulate the lower levels in polynomial time with queries to~$A$ and to a $\SigmakP{k-1}$ oracle and apply the argument of~\cite{HH90} to the top level, which gives the same with a $\SigmakP{k}$ oracle.

Our setting differs in two respects. First, our machines are not robust: they are unambiguous only on queries in the promise set and only with oracles consistent with the promise problem one level below. Second, our generator has no access to the promise oracles (it has only an $\NP$ oracle),
so the answers to promised queries are not available for checking a path. We therefore replace the learned answers by sets of candidate oblivious witnesses for every level and every query length. Relative to these sets, both answers of the oracles are certified: a positive answer by an accepting witness, and a negative answer by the rejection of all candidates. The generator looks for an input with two certified runs that give different outputs. By loose access, such a disagreement contains an incorrect answer to a query in the promise set; the uniqueness and the obliviousness of the witnesses on such queries take the place of the conflicts between accepting paths and let us trace it down the tower of oracles to a missing correct witness, which is added. Since the witnesses are oblivious, the number of rounds is bounded by the number of pairs (level, query length), and the sizes of the sets are bounded by descending induction on the level. Consequently, one $\NP$ oracle serves all levels at once, without the level-by-level induction of~\cite{ST07}, and the generator produces one advice string for all inputs of a given length instead of deciding a given input.

\paragraph{Dot-operators and single queries.}
The collapse in Corollary~\ref{cor:dot-collapse} uses the idea, applied in~\cite{HVW26} to prove that $\SOUP$ is closed under polynomial-time Turing reductions,
to test a candidate oblivious witness on a fixed input: since the outer witness is unique, the inner $\co\OUP$ predicate is false on one of two fixed pairs, so the oblivious witness refuting it can be identified without the actual input, which gives $\douexists\co\OUP=\OUP\cup\co\OUP$ (Theorem~\ref{th:dot}). In contrast, the non-oblivious dot-operator-based unambiguous hierarchy $\AUPH$ is infinite relative to an oracle~\cite{ST07}. 

To show that single queries to promise problems can be converted to language queries, we combine the ``input-tagging'' argument (a query~$q$ asked on input~$x$ is replaced by the pair~$(x,q)$, so that the oracle can check that $q$ is indeed asked on~$x$) of Cai, Hemachandra, and Vysko\v{c}~\cite[Proposition~2.6]{CHV93}, who showed that for deterministic machines, guarded nonadaptive access to $\pr\UP$ is not more powerful than access to $\UP$, with the observation that a single loose query of an unambiguous computation is guarded whenever its answer matters: otherwise, answering it in two ways would violate either correctness or independence of the computation result from the oracle's answer.

\paragraph{Organization of the paper.}
Section~\ref{sec:defs} contains definitions, notation, conventions, and known facts.
Section~\ref{sec:generators} contains preparations concerning generators with an $\NP$ oracle. The next sections contain the main results of the paper: Sections~\ref{sec:ouh} and~\ref{sec:loose} treat $\OUH$ and $\POUH$, respectively, and Section~\ref{sec:dot} considers the first levels of the hierarchy (the hierarchy defined by dot-operators, single queries, and the class $\OUP\cap\co\OUP$).
\section{Preliminaries}\label{sec:defs}
For rigorous definitions of standard notions of complexity theory, we refer to~\cite{AB09}.

\subsection{Conventions, injective pairing, promise problems, and oracle access}\label{subsec:conv}
All machines are polynomial-time unless stated otherwise; $n$ denotes the length of the input. In definitions based on polynomial-time predicates, all witnesses, random strings, and advice strings for inputs of length $n$ have prescribed lengths given by polynomials in $n$, and quantifiers range over strings of these lengths. Pairs and tuples are encoded in a length-regular way: if the lengths of the components are determined by $n$, the length of the encoding determines $n$ and the lengths of the components. A language $T$ is \emph{tally} if $T\subseteq\{1\}^*$; $\TALLY$ is the class of tally languages. We fix an injective pairing $\pi$ of a constant number of natural numbers such that $\pi$ is polynomially bounded in its arguments and $\pi$ and its inverse are computable in time polynomial in the unary lengths; we use it to encode tuples of numbers by tally strings. By $\lex$ we denote the lexicographic order on strings of the same length.

A \emph{promise problem} (introduced by Even, Selman, and Yacobi; see~\cite[Section~23.4]{AB09}) is a pair $\Pi=(\Pi_{\YES},\Pi_{\NO})$ of disjoint sets of strings; a language $L$ is identified with $(L,\overline L)$. We write $\Pi(x)=1$ for $x\in\Pi_{\YES}$ and $\Pi(x)=0$ for $x\in\Pi_{\NO}$. A language $O$ is \emph{consistent} with $\Pi$ if $\Pi_{\YES}\subseteq O$ and $O\cap\Pi_{\NO}=\emptyset$. For a class $\mathcal C$ of languages defined by machines, $\pr\mathcal C$ is the corresponding class of promise problems: the correctness requirements, as well as the semantic requirements such as the uniqueness of witnesses, are imposed only on the promise set $\Pi_{\YES}\cup\Pi_{\NO}$~\cite{HS15}. For a class $\mathcal D$ of promise problems, the \emph{class of languages in} $\mathcal D$ consists of the languages $L$ such that $(L,\overline L)\in\mathcal D$.

Our default oracle access mode to promise problems is \emph{loose}: queries outside the promise set are allowed and may be answered arbitrarily, and the computation must satisfy all requirements of its class for all such answers. These answers need not be consistent: the same query may receive different answers on different computation paths. We assume that a computation path never repeats a query. Then, for deterministic oracle machines, loose access is equivalent to the requirement of correctness for every language consistent with the oracle~\cite[Proposition~5.9]{CHV93}.
For unambiguous oracle machines, loose access is equivalent to the definition of~\cite[Definition~4]{HS15}: the computation must be correct for every consistent language, and its accepting witness, if any, must be the same for all consistent languages. (Indeed, the answers received on a single computation path are given by some consistent language; and if two consistent languages produced different accepting witnesses, then combining their answers on the two corresponding paths would produce two accepting paths.) We use this formulation below.

Accordingly, $\P^\Pi$ is the class of languages decided by deterministic polynomial-time oracle machines $M$ such that $M^O$ gives the correct answer for every language $O$ consistent with $\Pi$; an $\FP^\Pi$ computation may have several legal outputs (for different consistent languages $O$), and all of them must be correct but do not need to be the same string.
The subscript $\parallel$ (as in $\P^\Pi_\parallel$ or $\FP^\Pi_\parallel$) means that all queries are computed before any answer is received. For a class $\mathcal D$ of promise problems, $\P^{\mathcal D}=\bigcup_{\Pi\in\mathcal D}\P^\Pi$, and similarly for other oracle classes; $\mathcal C^{\mathcal D[1]}$ means that at most one query is made on every computation path.

\subsection{Unambiguous and input-oblivious classes}
We define the main classes interesting to us in their general promise relativized versions.
\begin{definition}[{$\UP$, $\OUP$~\cite{HVW26}, $\ONP$~\cite{FSW09}}]\label{def:classes}
Let $\Pi$ be a promise problem. A promise problem $\Gamma$ belongs to $\pr\UP^\Pi$ if there is a deterministic polynomial-time oracle machine $V$ such that for every $x\in\Gamma_{\YES}\cup\Gamma_{\NO}$, the set
\[
 W(x)=\{w : V^O(x,w)=1\}
\]
is the same for all languages $O$ consistent with $\Pi$, $|W(x)|=1$ for $x\in\Gamma_{\YES}$, and $W(x)=\emptyset$ for $x\in\Gamma_{\NO}$. If, moreover, there are strings $w_n$ such that $W(x)=\{w_{|x|}\}$ for every $x\in\Gamma_{\YES}$, then $\Gamma\in\pr\OUP^\Pi$, and $w_n$ are called \emph{oblivious unambiguous witnesses}. The problem $\Gamma$ belongs to $\pr\ONP^\Pi$ if there are a deterministic polynomial-time oracle machine $V$ and strings $w_n$ such that for every language $O$ consistent with $\Pi$, $V^O(x,w_{|x|})=1$ for every $x\in\Gamma_{\YES}$, and $V^O(x,w)=0$ for every $x\in\Gamma_{\NO}$ and every $w$. The classes $\UP^\Pi$, $\OUP^\Pi$, and $\ONP^\Pi$ consist of the languages in these classes; for a language oracle $\Pi$ (in particular, without an oracle) we obtain the usual definitions.
\end{definition}

\subsection{Input-oblivious unambiguous hierarchies}\label{subsec:hier}

\begin{definition}[input-oblivious unambiguous hierarchies]\label{def:hierarchies}
Define $\OUSigmaP{0}=\P$, $\pr\POUHK_0=\pr\P$, and, for $k\ge0$,
$
 \OUSigmaP{k+1}=\OUP^{\OUSigmaP{k}}
$
and
$
\pr\POUHK_{k+1}=\pr\OUP^{\pr\POUHK_k}
$
(recall that the access to promise oracles is loose). Let $\OUH=\bigcup_k\OUSigmaP{k}$ and $\pr\POUH=\bigcup_k\pr\POUHK_k$. The classes $\POUHK_k$ and $\POUH$ consist of the languages in the respective classes of promise problems.
\end{definition}

Note that $\P^{\pr\POUH}=\POUH$: a deterministic polynomial-time machine with loose access to $\Pi\in\pr\POUHK_k$ is an $\OUP^\Pi$ verifier with the empty witness, so its language belongs to $\POUHK_{k+1}$; the converse inclusion needs one query.

\subsection{Dot-operators}\label{subsec:dot}

\begin{definition}[{\cite[Definition~2.8]{HVW26}}]\label{def:dot}
Let $\mathcal C$ be a class of promise problems. A language $L$ belongs to $\douexists\mathcal C$ if there are $\Pi\in\mathcal C$, a polynomial $p$, and strings $w_n\in\bit^{p(n)}$ such that for every $n$ and every $x\in\bit^n$,
\begin{align*}
 x\in L&\ \Rightarrow\ (x,w_n)\in\Pi_{\YES}\text{ and }(x,w)\in\Pi_{\NO}\text{ for all }w\neq w_n,\\
 x\notin L&\ \Rightarrow\ (x,w)\in\Pi_{\NO}\text{ for all }w.
\end{align*}
If the witness $w_n$ may depend on $x$, we obtain the unique existential operator $\duexists$ (denoted $\exists!$ in~\cite{LR94,NR98,ST09}). If, instead, we drop the condition on the strings $w\neq w_n$, we obtain the oblivious existential operator $\doexists$~\cite{HVW26}, and if, moreover, the witness may depend on~$x$, the existential operator $\dexists$. The dual operator is $\douforall\mathcal C=\co\douexists\co\mathcal C$, where $\co$ exchanges $\Pi_{\YES}$ and $\Pi_{\NO}$. We allow the witness length to be identically zero (identifying $(x,\varepsilon)$ with $x$); thus the languages in $\mathcal C$ belong to $\douexists\mathcal C$.
\end{definition}

For $\douexists$ and $\duexists$, all pairs $(x,w)$ with $|w|=p(|x|)$ belong to the promise set of $\Pi$. Note that $\OUP=\douexists\P$ and, for every oracle $A$, $\OUP^A=\douexists\P^A$ (cf.\ \cite[Corollary~4.2]{HVW26}). The same holds for loose access to a promise problem~$\Pi$: $\OUP^\Pi=\douexists\P^\Pi$. Indeed, by Definition~\ref{def:classes}, the set of witnesses that an $\OUP^\Pi$ verifier accepts on an input~$x$ is the same for all languages consistent with~$\Pi$, so the accepted pairs $(x,w)$ form a language in~$\P^\Pi$; the converse is immediate.

\subsection{Generator--tester--decoder computations}\label{subsec:ogtd}

\begin{definition}[$\OGTD$, $\OUGTD$~\cite{HV26b}]\label{def:ogtd}
Let $\mathcal X$ be a class of promise problems. A \emph{generator} in $\FP^{\mathcal X}$ is a deterministic polynomial-time oracle transducer $G$ with an oracle $\Pi\in\mathcal X$ that outputs a string of length $p(n)$ on input $1^n$, for a polynomial $p$. Its set of \emph{legal outputs} is
\[
 \Out_G(n)=\{G^O(1^n): O\text{ is a language consistent with }\Pi\}.
\]
Let $\mathcal B$ and $\mathcal C$ be classes of languages. A language $L$ belongs to $\OGTD[\FP^{\mathcal X},\mathcal B,\mathcal C]$ if there are a generator $G\in\FP^{\mathcal X}$, a \emph{tester} $T\in\mathcal B$ (a language of pairs $(1^n,a)$), and a \emph{decoder} $D\in\mathcal C$ (a language of pairs $(x,a)$) such that, for every $n$,
\[
 \Out_G(n)\subseteq\Acc_T(n)\subseteq\Good_D(n),
\]
where $\Acc_T(n)=\{a:(1^n,a)\in T\}$ and $\Good_D(n)=\{a\in\bit^{p(n)}:\forall x\in\bit^n\ D(x,a)=L(x)\}$. The language $L$ belongs to $\OUGTD[\FP^{\mathcal X},\mathcal B,\mathcal C]$ if, moreover, $|\Acc_T(n)|=1$ for every $n$. The symbol $\ast$ in place of $\mathcal B$ means that no tester is required, that is, only $\Out_G(n)\subseteq\Good_D(n)$ is required.
Finally, $\OGTD[\mathcal X]=\OGTD[\FP^{\mathcal X},\ast,\P]$.
\end{definition}

\begin{proposition}[{\cite[Proposition~2.11]{HV26b}}]\label{prop:ogtd-facts}
Let $\Pi$ be a promise problem. Then $\OGTD[\Pi]\subseteq\P^{\Pi}\cap\Ppoly$.
\end{proposition}

We also use the notation of~\cite{KT94} for classes with bounded advice functions: for a class $\mathcal C$ of languages and a class $\mathcal F$ of advice functions, $\mathcal C/\mathcal F$ is the class of languages $L$ such that $x\in L\Leftrightarrow(x,h(1^{|x|}))\in B$ for some $B\in\mathcal C$ and $h\in\mathcal F$. %

\begin{definition}[$\sOprMA$, $\OUM$~\cite{HV26b}]\label{def:oum}
A promise problem $\Pi$ belongs to $\sOprMA$ if there are a polynomial-time predicate $V(x,y,r)$ and strings $y_n$ such that for every $x\in\bit^n$: if $x\in\Pi_{\YES}$, then $V(x,y_n,r)=1$ for all $r$; if $x\in\Pi_{\NO}$, then $\Pr_r[V(x,y,r)=1]<1/4$ for all $y$; and, for all $y$ and $r$, $V(x,y,r)=1$ implies $V(x,y_n,r)=1$. Define $\OUM=\OUGTD[\FP^{\sOprMA},\co\ONP,\P]$.
\end{definition}

Every language in $\ONP$ belongs to $\sOprMA$: take $V(x,y,r)=V'(x,y)$ for an $\ONP$ verifier $V'$~\cite{HV26b}. Hence $\FP^{\ONP}$ generators are $\FP^{\sOprMA}$ generators.

\subsection{Useful facts about computations with tally \texorpdfstring{$\NP$}{NP} oracles}\label{subsec:tally}
Queries to a tally oracle can be parallelized, because all of them can be computed during a preparation stage, as the (polynomial) range of all possible unary queries is known (cf., e.g.,~\cite{KT94}). Since tally $\NP$ languages are in $\ONP$ (cf.~\cite{FSW09,GM15}), one can formulate the following useful fact.

\begin{proposition}\label{prop:parallel}
$\P^{\TALLY\cap\NP}=\P^{\TALLY\cap\NP}_\parallel\subseteq\P^{\ONP}_\parallel$.
\end{proposition}

\begin{lemma}\label{lem:tally}
For a set $C\in\TALLY\cap\NP$, any tally set in $\NP^C$ belongs to $\P^{\TALLY\cap\NP}$.
\end{lemma}
\begin{proof}
Let $U\in\NP^C$ be a tally language, accepted by a nondeterministic polynomial-time oracle machine $N$ with the oracle $C$. If $U$ were in $\NP$, our deterministic machine would just query $U$. However, the machine $N$ uses the oracle $C$. We will get rid of this oracle by replacing it by an additional input to the $\NP$ machine (so our deterministic machine will query a different tally language $D$). We can compute the relevant portion of the truth table of $C$ deterministically using the oracle $C$. However, we cannot use this truth table as the additional input, because it is too large to embed into a tally language. Instead, this additional input will be a much shorter ``checksum'': the number of strings of relevant lengths in $C$ (which we can also compute using the oracle $C$).

Let $h(m)$ be the running time bound of $N$ on input $1^m$; it will bound the length of any query of $N$ on this input. Let $C_m$ contain all the strings in $C$ of length at most $h(m)$, and let $c_m=|C_m|$. Knowing $c_m$, an $\NP$ machine can guess $C_m$ (that is, $c_m$ numbers $k\le h(m)$) together with $\NP$ witnesses for the elements of $C_m$; it can then deterministically verify these witnesses to make sure the guessed strings are indeed in $C$. After that this machine can answer queries to $C$ itself.

Formally, consider the tally language $D$ of all strings $1^{\pi(m,c)}$ (recall that $\pi$ denotes an injective pairing) for which there are a $c$-size set $S$ of numbers $\le h(m)$, correct $\NP$ witnesses of $1^k\in C$ for each $k\in S$, and an accepting computation of $N$ on input $1^m$ in which the queries $1^k$ for $k\in S$ are answered positively and all other queries are answered negatively. Clearly, $D\in\NP$. For $c=c_m$ the only such set $S$ is $\{k\in\mathbb{N}:1^k\in C_m\}$: all elements of $S$ must have correct witnesses, and the checksum makes sure all elements of $C_m$ are listed. Hence the answers given to $N$ are those of $C$, so $1^{\pi(m,c_m)}\in D$ if and only if $1^m\in U$. Thus, to decide whether $1^m\in U$, a deterministic polynomial-time machine asks the oracle $C$ about $1^k$ for all $k\le h(m)$, which gives $c_m$, and then asks the oracle $D$ about $1^{\pi(m,c_m)}$. The tally oracles $C$ and $D$ can be combined into a single oracle in $\TALLY\cap\NP$ using $\pi$.
\end{proof}

\section{Generators with an \texorpdfstring{$\NP$}{NP} oracle}\label{sec:generators}
Our special focus is on $\OGTD[\NP]$, which, as we show below (Corollary~\ref{cor:popt}), is the class $\P/\OptP$ studied long ago by K\"obler and Thierauf~\cite{KT94}. 
But first we show that in this class the advice computed by the generator can be made ``canonical'' (that is, unambiguous) and testable in~$\co\ONP$.

We follow Krentel's proof~\cite[Theorem~3.1]{Kre88}: the correct sequence of answers of an $\NP$ oracle is the lexicographically largest sequence in which all positive answers are certified. We need to add the witnesses for the positive answers to the sequence, to make a candidate checkable in polynomial time, and to use that on the input~$1^n$ the largest feasible string~$a_n$ depends only on~$n$, to make it a common witness for the queries of the generator and a common refutation of all other strings; this gives the $\ONP$ oracle and the $\co\ONP$ tester.
\begin{lemma}\label{lem:canonical}
$
 \OGTD[\NP]=\OGTD[\ONP]=\OUGTD[\FP^{\ONP},\co\ONP,\P].
$
\end{lemma}
\begin{proof}
The containments $\OUGTD[\FP^{\ONP},\co\ONP,\P]\subseteq\OGTD[\ONP]\subseteq\OGTD[\NP]$ are trivial.
Let us replace the $\NP$ oracle of the generator by an $\ONP$ oracle and make the advice string unique by using the additional $\co\ONP$ tester.

Let $L\in\OGTD[A]$ with a generator $G\in\FP^A$ and a decoder $D\in\P$, where $A\in\NP$ has a polynomial-time verifier $U$. Our new generator will output something the old advice can be reconstructed from: the sequence $a$ of oracle answers for the queries of $G(1^n)$, followed by the list of the corresponding witnesses for the positive answers (padded by zeroes to a fixed length sufficient in any possible case). It can be verified in polynomial time by simulating the run of $G$ and verifying the witnesses with $U$. The only thing that cannot be verified is the negative answers. Therefore, we will require the lexicographically largest such string $a_n$: if we find it, then, in particular, all answers will be correct; also the largest string is guaranteed to be unique. 
The property that a formally correct candidate string is the largest can be tested in $\co\ONP$: the correct answer $a_n$ is the oblivious witness.

Let $R(1^n,a)$ be the polynomial-time deterministic procedure that checks that $a$ is consistent with the run of $G(1^n)$ using the answers from $a$, and that all the witnesses are correct.
We define our $\co\ONP$ tester as
\[
 T=\{(1^n,a): R(1^n,a)\land\forall b\gtlex a\ \neg R(1^n,b)\},
\]
thus it accepts solely $a=a_n$.
Our new generator will use the oracle $Q=\{(1^n,c):\exists a\gelex c\ R(1^n,a)\}\in\ONP$ (where $a_n$ plays again the role of the oblivious witness). It will find $a_n$ by prefix search using this oracle. Namely, having found the first $j-1$ bits $b_1\ldots b_{j-1}$ of $a_n$, it asks whether $(1^n,c)\in Q$ for the string $c=b_1\ldots b_{j-1}10\ldots0$ of the same length as $a_n$. Since $a_n$ is the largest string $a$ with $R(1^n,a)$, we have $(1^n,c)\in Q$ if and only if $a_n\gelex c$, that is, if and only if the $j$-th bit of $a_n$ is~$1$.

The new decoder will reconstruct the old advice from $a_n$ and run the old decoder. 
\end{proof}

\begin{corollary}\label{cor:popt}$\OGTD[\NP]=\P/\OptP$.\end{corollary}
\begin{proof}This is a corollary of the proof rather than of the formulation of Lemma~\ref{lem:canonical}. We generate $a_n$ as the maximum of all strings $a$ with $R(1^n,a)$, which gives an $\OptP$ function computing $a_n$ on input $1^n$. In the other direction, every function in $\OptP$ can be computed in $\FP^{\NP}$ by binary search~\cite{Kre88}. 
\end{proof}

\section{The oblivious unambiguous hierarchy}\label{sec:ouh}
We show that the oblivious unambiguous hierarchy based on languages is contained in the second level of the oblivious hierarchy; moreover, its languages can be computed with nonadaptive queries to tally $\NP$ oracles.

\begin{theorem}\label{th:ouh}
$\OUH\subseteq\P^{\TALLY\cap\NP}$.
\end{theorem}
\begin{proof}
It suffices to show that $\OUP^{\P^{\TALLY\cap\NP}}\subseteq\P^{\TALLY\cap\NP}$ and use it repeatedly. 

Let $L\in\OUP^A$ with $A\in\P^T$ for some $T\in\TALLY\cap\NP$, and let $V$ be a polynomial-time verifier for $L$ with oblivious unambiguous witnesses $w_n$ of length $p(n)$. Let $V'$ be a version of $V$ that simulates the oracle $A$ itself using its $\P^T$ algorithm (so $V'$ uses $T$ as an oracle).

Since $w_n$ is unique, we can compute it bit by bit (obliviously of the input) using queries to the following tally language
\[
 W=\{1^{\pi(n,i)}:1\le i\le p(n)\text{ and }V'^T(x,w)=1\text{ for some }x\in\bit^n\text{ and }w\text{ with }w_i=1\}.
\]
If $L\cap\bit^n\neq\emptyset$, then $w_n=W(1^{\pi(n,1)})\ldots W(1^{\pi(n,p(n))})$. If $L\cap\bit^n=\emptyset$, then this string is $0^{p(n)}$, which is as good as any other string in this case, so we can assume w.l.o.g. $w_n=0^{p(n)}$. 
It is readily seen from that definition of $W$ that $W\in\NP^T$. By
Lemma~\ref{lem:tally}, 
$W$ can be computed in deterministic polynomial time with some oracle $O\in{\TALLY\cap\NP}$.

Now, $L$ can be decided by first computing $w_n$ and then simulating $V'^T$ on the actual input with this witness. The resulting deterministic algorithm uses tally $\NP$ oracles $T$ and $O$, which can be combined in a single oracle using $\pi$.
\end{proof}

\begin{corollary}\label{cor:ouh}
$\OUH\subseteq\P^{\ONP}_\parallel$. 
\end{corollary}
\begin{proof}
By Theorem~\ref{th:ouh} and Proposition~\ref{prop:parallel}.
\end{proof}

\section{The loose oblivious unambiguous hierarchy}\label{sec:loose}
The proof of Theorem~\ref{th:ouh} uses the fact that the oracle is a language: for a promise oracle, the search for an accepting witness may stumble upon a witness accepted on an input outside the promise. We will now overcome this difficulty at the expense of switching to the possibly larger class $\OGTD[\NP]$, whose generator makes adaptive $\NP$ (essentially, $\ONP$) queries (a nonadaptive one would give $\P^{\TALLY\cap\NP}$: its queries $q_{n,1},q_{n,2},\ldots$ on~$1^n$ do not depend on the oracle answers, so its oracle $A\in\NP$ can be replaced by the tally language $\{1^{\pi(n,j)}:q_{n,j}\in A\}$, which is in $\NP$%
), and accordingly our deterministic simulation with an $\ONP$ oracle becomes adaptive (compare Corollaries~\ref{cor:ouh} and~\ref{cor:loose}).

\begin{theorem}\label{th:loose}
$\POUH\subseteq\OGTD[\NP]$.
\end{theorem}
\begin{proof}
Recall that $\POUH=\P^{\pr\POUH}$.
Let $L\in\P^{\Pi_d}$, where $\Pi_d\in\pr\OUP^{\Pi_{d-1}}$, and let $M$ be the corresponding polynomial-time deterministic oracle machine.
The problem $\Pi_d$ is defined inductively through promise problems at previous levels,
that is, there are $\Pi_0\in\pr\P$ and, for $1\le i\le d$, $\Pi_i\in\pr\OUP^{\Pi_{i-1}}$, and there are oracle polynomial-time deterministic machines (verifiers) $V_i$ for $\Pi_i$ with oblivious unambiguous witnesses $w_{i,m}$. 
All the machines must work correctly for any answers to the oracle queries outside the promise set, in particular, for queries answered by any languages $A_i$ consistent with $\Pi_i$. Let us define specific such languages.  Fix a polynomial-time language $A_0$ consistent with $\Pi_0$, and define
\[
 A_i=\{q:\exists w\ V_i^{A_{i-1}}(q,w)=1\}\qquad(1\le i\le d).
\]
By induction on $i$, the language $A_i$ is consistent with $\Pi_i$; in particular, $M^{A_d}$ must decide $L$. Note that we do not replace loose access with access via a consistent language in the definition of $\POUH$; these languages are used only in the analysis.

We will replace these languages $A_i$ with different languages $B_i$ that have the same effect. Then queries to $B_d$ will be handled within $\OGTD[\NP]$.

\medskip\noindent
\emph{The advice string.} 
Fix the input length $n$, and let $h=h(n)$ be a polynomial bounding the lengths of all queries and witnesses and the running times of $M$ and of all $V_i$ in the simulation of $M$ on inputs of length $n$, for all witnesses and all oracle answers. The advice that our $\FP^{\NP}$ generator will compute will contain all the witnesses we might use on all the lengths we can reach while running $M$ on an input of length $n$. It will be a gradually constructed family $S=(S_{i,m})_{1\le i\le d,\,0\le m\le h}$ of sets of witnesses, where $S_{i,m}$ consists of candidate witnesses for $V_i$ on queries of length $m$; we put $S_{i,m}=\emptyset$ for $m>h$.

\medskip\noindent
\emph{Restricting $V_i$ to work with a subset of candidate witnesses; certified runs.}
A \emph{run} of an oracle machine is a protocol of its behavior including the answers it received to its oracle queries (including those missing the intended promise set). We will call a run of $M(x)$ or of a $V_i(q,u)$ \emph{certified} (w.r.t.\ specific $S$) if we can provide certificates for the answers to all its oracle queries in the following sense. We define certificates for the answers to level-$i$ queries, i.e., queries to $\Pi_i$. At level~$0$, an answer to $q$ is certified if it simply equals $A_0(q)$ (recall that $A_0\in\P$). For $i\ge1$, a certificate for a positive answer to $q$ consists of a witness $w$ and an accepting run of $V_i(q,w)$ together with certificates for all oracle answers in this run; a certificate for a negative answer consists, for every $s\in S_{i,|q|}$, of a rejecting run of $V_i(q,s)$ together with certificates for all oracle answers in these runs (if $S_{i,|q|}=\emptyset$, the empty certificate suffices for a negative answer). 
Note that, for a constant level $d$, certificates have size polynomial in $n$ and the total number of witnesses in~$S$, and they can be verified in polynomial time given $1^n$ and $S$. 

\medskip\noindent
\emph{New languages $B_i$ and the decoder.}
If one uses the languages $A_i$ to give the oracle answers, they are all certified w.r.t.\ every~$S$: 
by induction on $i$, for $A_i(q)=1$ use an accepting witness, and for $A_i(q)=0$ use the rejecting runs of $V_i^{A_{i-1}}(q,s)$ for $s\in S_{i,|q|}$. Define $B_0=A_0$ and
\[
 B_i=\{q:\exists s\in S_{i,|q|}\ V_i^{B_{i-1}}(q,s)=1\}\qquad(1\le i\le d);
\]
the answers of $B_i$ are certified similarly. The languages $B_i$ need not be consistent with $\Pi_i$ (their definition is restricted to witnesses in $S$), but, given $S$, $B_d(q)$ is computable in time polynomial in $n$ and the total number of witnesses in~$S$.
So we define our decoder as $D(x,S)=M^{B_d}(x)$, and it can be implemented in deterministic polynomial time.

\medskip\noindent
\emph{The generator.} Start with $S=\emptyset$ and repeat the following round. 

Ask the $\NP$ oracle whether $M$ has two certified runs w.r.t. $S$ with different outputs for some $x\in\bit^n$. If it has none, the process finishes and the generator outputs $S$, which is already good enough: since both runs $M^{A_d}(x)$ and $M^{B_d}(x)$ are certified, we have $D(x,S)=M^{B_d}(x)=M^{A_d}(x)=L(x)$ for every $x\in\bit^n$. 

Otherwise, find this input $x$ and two such runs together with all their certificates by repeatedly querying their prefixes with the $\NP$ oracle, and find the problem (which should be a missing correct witness --- and indeed it will be found, as we show below). 
Namely, consider the two runs and check them recursively, examining the certificates of all their answers. For every negative answer to a level-$i$ query $q$ with $i\ge1$, ask the $\NP$ oracle whether $q$ has a certificate for a positive answer (with respect to the same $S$); if it does, find one, schedule its witness $w$ for the addition to $S_{i,|q|}$, and check it as well. After this recursive process finishes, add all scheduled witnesses to $S$, and repeat the round.

\medskip
We need to check how many rounds we need. To prove that a polynomial number of rounds suffices, it is enough to prove that each round adds some correct witness $w_{i,m}$ to some set $S_{i,m}$ --- since we deal with unambiguous oblivious computations, there are at most $d(h+1)$ correct witnesses. Incorrect witnesses can be added too: a certificate for a positive answer to a level-$i$ query~$q$ may contain an incorrect witness if $q$ is outside the promise set of~$\Pi_i$ (where $V_i$ may accept several witnesses) or if the certificate contains an incorrect answer to a promised query at a lower level (as in the third case of the claim below), which is possible only while some correct witness at a lower level is missing from~$S$. Such witnesses stay in~$S$, but they do not harm our computations (we verify everything) and do not take too much time (the number of witnesses that a round adds at level~$i$ is bounded by a polynomial in~$h$ and the sizes of the sets at the levels above~$i$: every level-$j$ query checked in the round leads to checking at most $h(\max_m|S_{j,m}|+1)$ level-$(j-1)$ queries); hence by induction the sizes of all sets stay polynomial during all the rounds.
In particular, the length we need to pad $S$ to is polynomially bounded and the generator's running time is polynomial.

\medskip\noindent
\emph{Remaining claim: each round adds at least one correct witness.}
Call an answer \emph{wrong} if it is an incorrect answer to a query in the promise set of the respective $\Pi_i$. We claim that every round adds to some $S_{i,m}$ the witness $w_{i,m}$ that was not there before. One of the two runs of $M$ has a wrong output; hence it contains a wrong answer, since otherwise it would be the run of $M^O(x)$ for some language $O$ consistent with $\Pi_d$. Similarly, a run of $V_i$ with all answers correct on promised queries is a run of $V_i^O$ for a language $O$ consistent with $\Pi_{i-1}$. Consider a wrong answer to a level-$i$ query $q$ discovered during the round ($i\ge1$, because at level 0 all certified answers are always correct):
\begin{itemize}
\item If the answer is positive, then $q\in\Pi_{i,\NO}$, and the certificate contains an accepting run of $V_i(q,w)$, which must contain a wrong answer at level $i-1$.
\item If the answer is negative and $w_{i,|q|}\in S_{i,|q|}$, then $q\in\Pi_{i,\YES}$, and the certificate contains a rejecting run of $V_i(q,w_{i,|q|})$, which must contain a wrong answer at level $i-1$.
\item If the answer is negative and $w_{i,|q|}\notin S_{i,|q|}$, then a certificate for the positive answer exists (as $A_i(q)=1$), and the generator finds one with some witness $w$. If $w=w_{i,|q|}$, the claimed addition is made. Otherwise, by the uniqueness of the witness for $q\in\Pi_{i,\YES}$, the accepting run of $V_i(q,w)$ in this certificate contains a wrong answer at level $i-1$.\qedhere
\end{itemize}

\end{proof}

\begin{remark}\label{rem:hh}
The generator follows the strategy of the simulation of robustly categorical machines by Hartmanis and Hemachandra~\cite[Theorem~2.1]{HH90}:
the sets $S_{i,m}$ are a more elaborate equivalent of the oracle answers learned there.
\end{remark}

\begin{corollary}\label{cor:loose}
$\POUH\subseteq\OUM$ and $\POUH\subseteq\P^{\ONP}$.
\end{corollary}
\begin{proof}
Theorem~\ref{th:loose} puts $\POUH$ into $\OGTD[\NP]$. 
Lemma~\ref{lem:canonical} proves that this class equals $\OGTD[\ONP]$
(thus putting it into $\P^{\ONP}$, cf.\ Proposition~\ref{prop:ogtd-facts}) and
strengthens it to the class
$
 \OUGTD[\FP^{\ONP},\co\ONP,\P]
$
(thus putting it into $\OUM$).
\end{proof}

\section{The first levels of the hierarchy}\label{sec:dot}

In this section we sort out what happens with even smaller classes.
We check what happens for even more restricted constructions of the hierarchy.
One can ask what happens if we attempt to define an oblivious unambiguous hierarchy 
using dot-operators --- we will show that it surprisingly collapses.
One can ask about applying our dot-operator to promise problems ---
it does not change the class.
Also dot-operators are somewhat similar to single oracle queries --- we will show that
for such queries a promise oracle can be replaced in our context by a language
(in contrast to what we know --- or do not know --- about promise $\RP$ and promise $\ONP$ queries).
Finally, we remark that in an even smaller class $\OUP\cap\co\OUP$
the witness can in fact be made the same for both parts, making this class equal to the class $\SOUP$
recently defined by \cite{HVW26}.

\subsection{A hierarchy defined via dot-operators collapses}
We start with two simple properties of the operator $\douexists$ (Definition~\ref{def:dot}) and then go straight to our first collapse.

\begin{proposition}\label{prop:dot}\phantom{.}
\begin{enumerate}
\item $\douexists\pr\mathcal C=\douexists\mathcal C$ for $\mathcal C\in\{\P,\OUP,\co\OUP\}$.
\item $\douexists\OUP=\OUP$.
\end{enumerate}
\end{proposition}
\begin{proof}
1. Restricting the machine for~$\Pi$ to the pairs $(x,w)$ with $|w|=p(|x|)$, which are all in the promise set, gives a language in~$\mathcal C$ that can replace~$\Pi$.

2. Merge the two witnesses into one (taking care of possible length clashes). 
\end{proof}

\begin{theorem}\label{th:dot}
$\douexists\co\OUP=\OUP\cup\co\OUP$.
\end{theorem}
\begin{proof}
Zero witness length gives $\co\OUP\subseteq\douexists\co\OUP$, and $\OUP=\douexists\P\subseteq\douexists\co\OUP$. Conversely, let $L\in\douexists R$, where $R\in\co\OUP$, and let $V$ be an oblivious unambiguous verifier for $\overline R$ with witnesses $v_N$. If $p\equiv0$, then $L\in\co\OUP$. Otherwise, $p(n)\ge1$ for all but finitely many $n$, and since finitely many input lengths can be handled separately, we may assume $p(n)\ge1$ for all $n$. At most one $w$ satisfies $(0^n,w)\in R$; hence at least one of the pairs $(0^n,0^{p(n)})$ and $(0^n,1^{p(n)})$ belongs to $\overline R$, and $v_N$, where $N$ is the length of these pairs, is the only string $v$ such that
\[
 V((0^n,0^{p(n)}),v)=1\ \lor\ V((0^n,1^{p(n)}),v)=1
\]
(testing a candidate oblivious witness on fixed inputs). On input $x\in\bit^n$, accept a pair $(w,v)$ if and only if this condition holds and $V((x,w),v)=0$. Then $(w_n,v_N)$ is the only accepted pair for $x\in L$, and no pair is accepted for $x\notin L$. Thus $L\in\OUP$.
\end{proof}

\begin{corollary}\label{cor:dot-collapse}
The hierarchy obtained from $\P$ by alternating applications of $\douexists$ and $\douforall$ collapses to its second level $\douexists\co\OUP=\douforall\OUP=\OUP\cup\co\OUP$: this class is closed under $\douexists$ and $\douforall$.
\end{corollary}
\begin{proof}
Clearly, $\douexists(\mathcal C\cup\mathcal D)=\douexists\mathcal C\cup\douexists\mathcal D$ for all classes $\mathcal C$ and $\mathcal D$. Apply Proposition~\ref{prop:dot}(2) and Theorem~\ref{th:dot} and take complements.
\end{proof}

This is in contrast with the polynomial hierarchy and with the oblivious hierarchy that give the same classes when defined via oracles and via dot-operators (and are not known to collapse), as well as with the non-oblivious analogue $\AUPH$ of our collapsed unambiguous dot-operator hierarchy, which is infinite relative to an oracle~\cite{ST07}. Our collapse uses both obliviousness and unambiguity substantially: since the outer witness is unique, the inner $\co\OUP$ predicate is false on one of the fixed pairs $(0^n,0^{p(n)})$ and $(0^n,1^{p(n)})$, and since the witness refuting this predicate is oblivious, it can be tested on that pair.

\subsection{The single query case}
For deterministic machines, Cai, Hemachandra, and Vysko\v{c}~\cite[Proposition~2.6]{CHV93} observed that guarded nonadaptive access to $\pr\UP$ gives no more power than access to $\UP$: tagging every query with the input yields a language in $\UP$. We show that a single loose query of an unambiguous computation is guarded whenever its answer matters, which makes the tagging argument applicable.

\begin{lemma}\label{lem:one-query}
Let $\Pi$ be a promise problem, and let $V$ be a polynomial-time oracle verifier that makes at most one query $q(x,w)$ on every input $(x,w)$ and witnesses $L\in\UP^\Pi$. Let $T$ be the set of pairs $(x,w)$ such that the output of $V(x,w)$ depends on the answer to $q(x,w)$. Then $T\in\P$, and $q(x,w)\in\Pi_{\YES}\cup\Pi_{\NO}$ for every $(x,w)\in T$.
\end{lemma}
\begin{proof}
To decide $T$, run $V(x,w)$ with both answers. Let $(x,w)\in T$, and suppose that $q=q(x,w)$ is outside the promise. Take languages $O_0$ and $O_1$ consistent with $\Pi$ that differ only on $q$. Then $V^{O_b}(x,w)=1$ for exactly one $b$. If $x\notin L$, this contradicts the correctness of $V^{O_b}$. If $x\in L$, then $w$ is the accepting witness of $V^{O_b}$ on $x$; since the accepting witness does not depend on the consistent language (Definition~\ref{def:classes}), $V^{O_{1-b}}(x,w)=1$, a contradiction.
\end{proof}

\begin{theorem}\label{th:one-query}
$\UP^{\pr\UP[1]}=\UP^{\UP[1]}$ and $\OUP^{\pr\OUP[1]}=\OUP^{\OUP[1]}$.
\end{theorem}
\begin{proof}
The containments ``$\supseteq$'' are trivial. Let $V$ and $\Pi$ be as in Lemma~\ref{lem:one-query}, where $V$ witnesses $L\in\OUP^\Pi$ and $\Pi\in\pr\OUP$ via a verifier $U$ (the case of $\UP$ is similar and simpler). Let $\langle x,w\rangle$ denote the pair $(x,w)$ padded so that its length determines both $|x|$ and $|q(x,w)|$, and let
\[
 A=\{\langle x,w\rangle:(x,w)\in T\text{ and }q(x,w)\in\Pi_{\YES}\}.
\]
Then $A\in\OUP$: on input $\langle x,w\rangle$, reject if $(x,w)\notin T$, and otherwise run $U$ on $q(x,w)$, which is in the promise set; all members of $A$ of the same length have queries of the same length and hence the same witness. The new verifier computes $V(x,w)$ without the oracle if $(x,w)\notin T$, and otherwise answers the query $q(x,w)$ by querying $\langle x,w\rangle\in A$. On every input $(x,w)$, it gives the same result as $V^O(x,w)$ for a fixed language $O$ consistent with $\Pi$; thus it witnesses $L\in\OUP^A$ with the same witnesses.
\end{proof}

For comparison, a single loose query to $\pr\RP$ gives $\RP^{\pr\RP[1]}=\BPP$ (by the proof of~\cite[Theorem~1]{BF99}), whereas $\P=\RP\neq\BPP$ relative to an oracle~\cite[Theorem~3]{BF99}. Similarly, a single loose query to $\pr\ONP$ gives $\ONP^{\pr\ONP[1]}=\doexists\pr\,\co\ONP=\OtP$
\cite[Theorem~4.7]{HVW26}; whereas $\ONP^{\ONP[1]}=\ONP^{\ONP}=\doexists\co\ONP$~\cite[Theorem~4.4]{HVW26}, and it is open whether these two classes differ.

\subsection{A note on \texorpdfstring{$\OUP\cap\co\OUP$}{OUP ∩ coOUP}}
It turns out that $\OUP\cap\co\OUP$ coincides with the class $\SOUP$~\cite[Definition~2.7]{HVW26} of languages $L$ such that $L$ and $\overline L$ have oblivious unambiguous verifiers with the same oblivious unambiguous witnesses. Thus it inherits the polynomial-time Turing closure property from $\SOUP$; it also follows that it is self-low.

\begin{proposition}\label{prop:soup}
$\SOUP=\OUP\cap\co\OUP$. Consequently, $\OUP^{\SOUP}=\OUP$ and $\SOUP^{\SOUP}=\SOUP$.
\end{proposition}
\begin{proof}
The inclusion $\SOUP\subseteq\OUP\cap\co\OUP$ is immediate. For the converse, let $Y$ and $Z$ be oblivious unambiguous verifiers for $L$ and $\overline L$, respectively, with (w.l.o.g.) equally long witnesses. On input $x\in\bit^n$, guess a bit $b$ and a string $w$ and check
\[
 (b=1\land Y(0^n,w)=1)\ \lor\ (b=0\land Z(0^n,w)=1).
\]
Since $0^n$ belongs either to $L$ or to~$\overline L$, exactly one pair passes this check. %
If $b=1$, its second component is the oblivious witness for $Y$, so $Y(x,w)=L(x)$. If $b=0$, similarly $1-Z(x,w)=L(x)$. In any case, this pair $(b,w)$ certifies either answer.

Let $A\in\SOUP$. Since $\P^A\subseteq\SOUP$ (as $\SOUP$ is closed under polynomial-time Turing reductions~\cite{HVW26}), Proposition~\ref{prop:dot}(2) gives $\OUP^A=\douexists\P^A\subseteq\douexists\OUP=\OUP$; hence $\co\OUP^A\subseteq\co\OUP$ and $\SOUP^A\subseteq\OUP^A\cap\co\OUP^A\subseteq\SOUP$.
\end{proof}

The class $\SOUP$ is an unambiguous version of $\YP^*$~\cite{HVW26}, and $\OUP\cap\co\OUP$ is an unambiguous version of $\YP=\ONP\cap\co\ONP$~\cite{AD14}; the proof of Proposition~\ref{prop:soup} relies on uniqueness and does not give $\YP=\YP^*$.

\section*{Acknowledgement}
The author is indebted to Ilya Volkovich and Ryan Williams, with whom he started
the study of unambiguous oblivious classes \cite{HVW26,HV26b}.

\end{document}